# ITERATIVE METHOD FOR IMPROVEMENT OF CODING AND DECRYPTION

Natasa Zivic

Institute for Data Communication Systems, University of Siegen, Siegen, Germany
natasa.zivic@uni-siegen.de

## ABSTRACT

*Cryptographic check values (digital signatures, MACs and H-MACs) are useful only if they are free of errors. For that reason all of errors in cryptographic check values should be corrected after the transmission over a noisy channel before their verification is performed. Soft Input Decryption is a method of combining SISO convolutional decoding and decrypting of cryptographic check values to improve the correction of errors in themselves. If Soft Input Decryption is successful, i.e. all wrong bit of a cryptographic check value are corrected, these bit are sent as feedback information to the channel decoder for a next iteration. The bit of the next iteration are corrected by channel decoding followed by another Soft Input Decryption.*

*Iterative Soft Input Decryption uses interleaved blocks. If one block can be corrected by Soft Input Decryption, the decoding of the interleaved block is improved (serial scheme). If Soft Input Decryption is applied on both blocks and one of the blocks can be corrected, the corrected block is used for an improved decoding of the other block (parallel scheme). Both schemes show significant coding gains compared to convolutional decoding without iterative Soft Input Decryption.*

.

## 1. INTRODUCTION

Modern communication systems use an encryptor and decryptor as standard components. Feedback is also used between the elements of the receiver [1], in order to improve demodulation and decoding results (Fig. 1). Unfortunately, the presence of a decryptor between source and channel decoder interupts and disables cooperation between these two decoding elements, which is known as joint source-channel decoding [2, 3]. For that reason two feedback loops instead of one should be realized: one from the source decoder to the decryptor and the other one from the decryptor to the channel decoder.

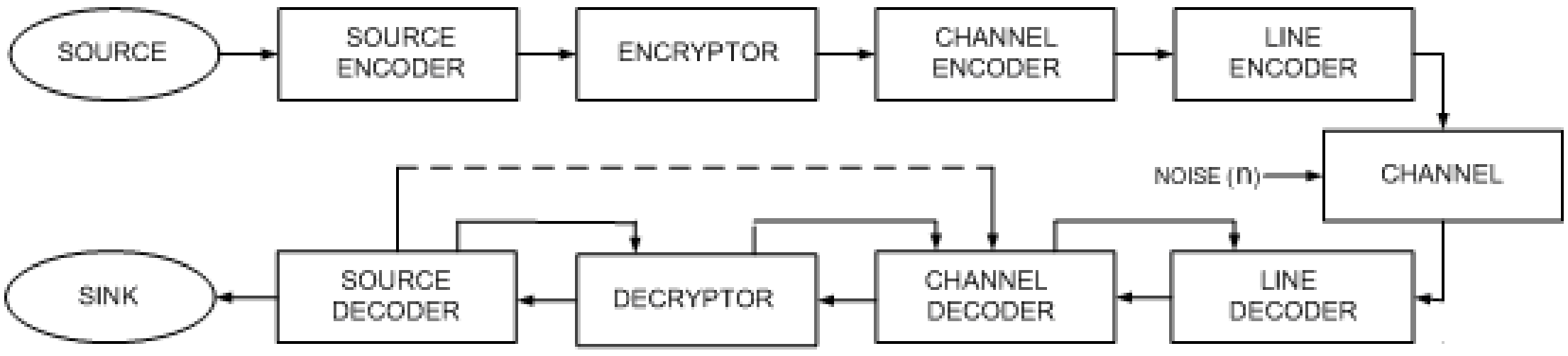

Fig.1 – Communications system with en- decryptor using feedback loops

The main problem by using cryptographic elements is that they need errorless input: if only one bit of the input of the decryptor is wrong, in average 50 % of its output bit are wrong and the received information is useless. In very noisy enviroments, as by wireless or satellite communications for example, errorless decoding is impossible and errors at the input of decryptor are present very often.

A possible solution for improved decoding applies a cooperation between coding and cryptography (chapter 2). Using the Soft Input Decryption method which is presented in chapter 3, many errors after decoding can be corrected. Soft Input Decryption is a combination of SISO convolutional channel decoding and decrypting.

Soft Input Decryption with feedback (chapter 4) includes the feedback between the decryptor and SISO convolutional decoder (see Fig. 1). The feedback enables the correction of bit decoded by a SISO decoder, using bit which have been previously corrected by Soft Input Decryption.

Iterative Soft Input Decryption is a method of Soft Input Decryption with feedback, which is extended by another Soft Input Decryption. Iterative Soft Input Decryption with is analyzed in chapter 5, using two strategies: serial and parallel. The results of the simulations are presented in chapter 6.

Chapter 7 gives an overview of the results and suggestions for the expansion of Iterative Soft Input Decryption to Turbo Soft Input Decryption, using logical analogy to turbo decoding [4].

As this area of telecommunications is relatively new, there are no much publications which examine the sinergy of cryptography and channel coding.

## 2. Cooperation between Channel Coding and Cryptography

This paper is based on the idea to use the soft output (reliability values or *L*-values) of SISO (Soft Input Soft Output) channel decoding to correct the input of inverse cryptographic mechanisms (decrypting) providing cryptographic redundancy. The channel code can be considered as an inner code and the output of the cryptographic mechanism as an outer code (Fig. 2). Cryptographic mechanisms are used for security reasons for the recognition of modifications by errors or different types of attacks.

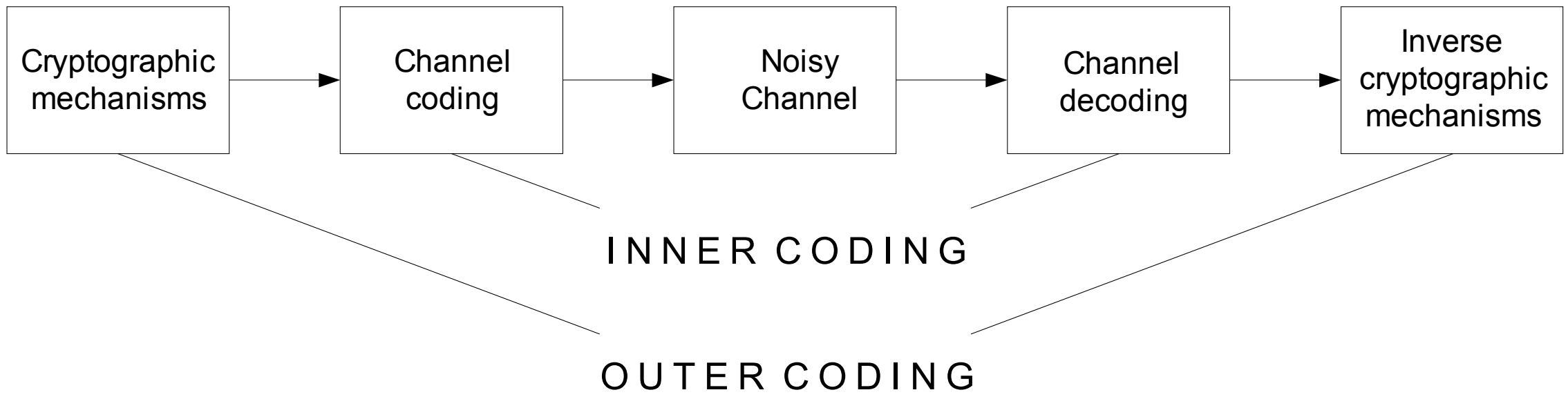

Fig. 2 Representation of channel coding and cryptographic mechanisms as inner and outer codes

Soft output values are information about decoded bits, which are used in today's most efficient decoders: turbo decoders. In this work $L$-values are used in a different way: as information to the next following entity – the decrypting mechanism.

Digital signatures (digital signatures with appendix [5] and digital signatures giving message recovery [6]), MACs (Message Authentication Codes [7]) and H-MACs (Hashed Message Authentication Codes [8]) are used in this paper as cryptographic check values because all of them are applied in practice and they have different lengths, which result in different coding gains. If the correction of a cryptographic check value is successful, feedback to the SISO channel decoder is accomplished for its improvement.

The concatenation of codes, presented as an outer and inner code was already devised by Forney in 1966 [9]. In literature, it is known as concatenated codes [10], general concatenated codes [11] or codes of a superchannel [12]. A general schema of a communication system using concatenated codes is presented in Fig. 3:

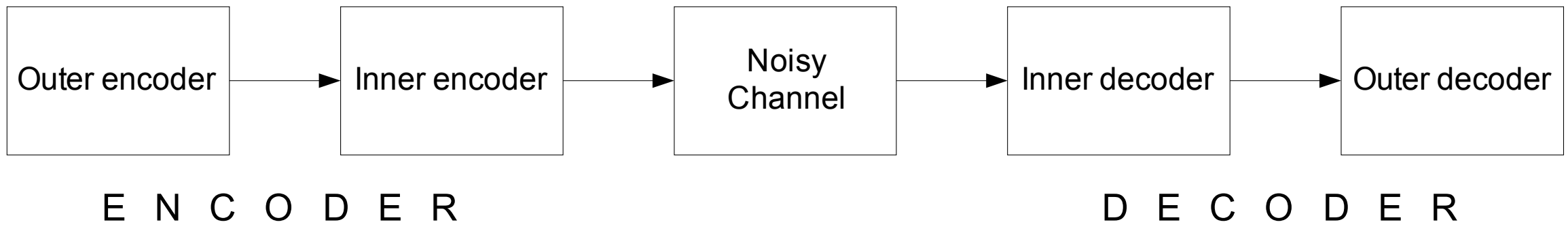

Fig. 3 Communication system using concatenated codes

In most cases a convolutional code is used as an inner code in combination with a Reed Solomon code or another convolutional code as an outer code. Such a type of concatenated codes can be compared to the combination of codes investigated in this work (Fig. 2). Two good characteristics are the result of such a concatenated schema: good error performance because of the use of the SISO principle and good security performance as result of the use of the cryptographic mechanisms.

## 3. Soft Input Decryption

A requirement for Soft Input Decryption is the usage of Soft Input – Soft Output (SISO) convolutional decoding. The output values of the SISO decoder ($L$-values) are used as information for the decryptor of Soft Input Decryption: a lower $|L|$-value indicates a higher probability that the decoded bit is wrong (if $|L| = 0$, the probability is 0.5), while a higher $|L|$-value indicates a lower probability that the decoded bit is wrong (if $|L| = \infty$, the bit is correctly decoded).

The Soft Input Decryption algorithm (Fig.4) works as follows [13]: if the verification of a cryptographic check value is negative, the soft output of the SISO decoder is analyzed and the bits with the lowest $|L|$-values are flipped (XOR 1) [14]; then the decryptor repeats the verification process and proves the result of the verification again. If the verification is again negative, bits with another combination of the lowest $|L|$-values are changed. This testing process is finished when the verification is successful or the provided resources are consumed e.g. maximal number of tests.

A sequence of bits of the output of the SISO channel decoder forms a so called SID block (Soft Input Decryption block). A SID block contains a digital signature (digital signature with appendix or digital signature giving message recovery), a message with its MAC, or a message with its H-MAC.

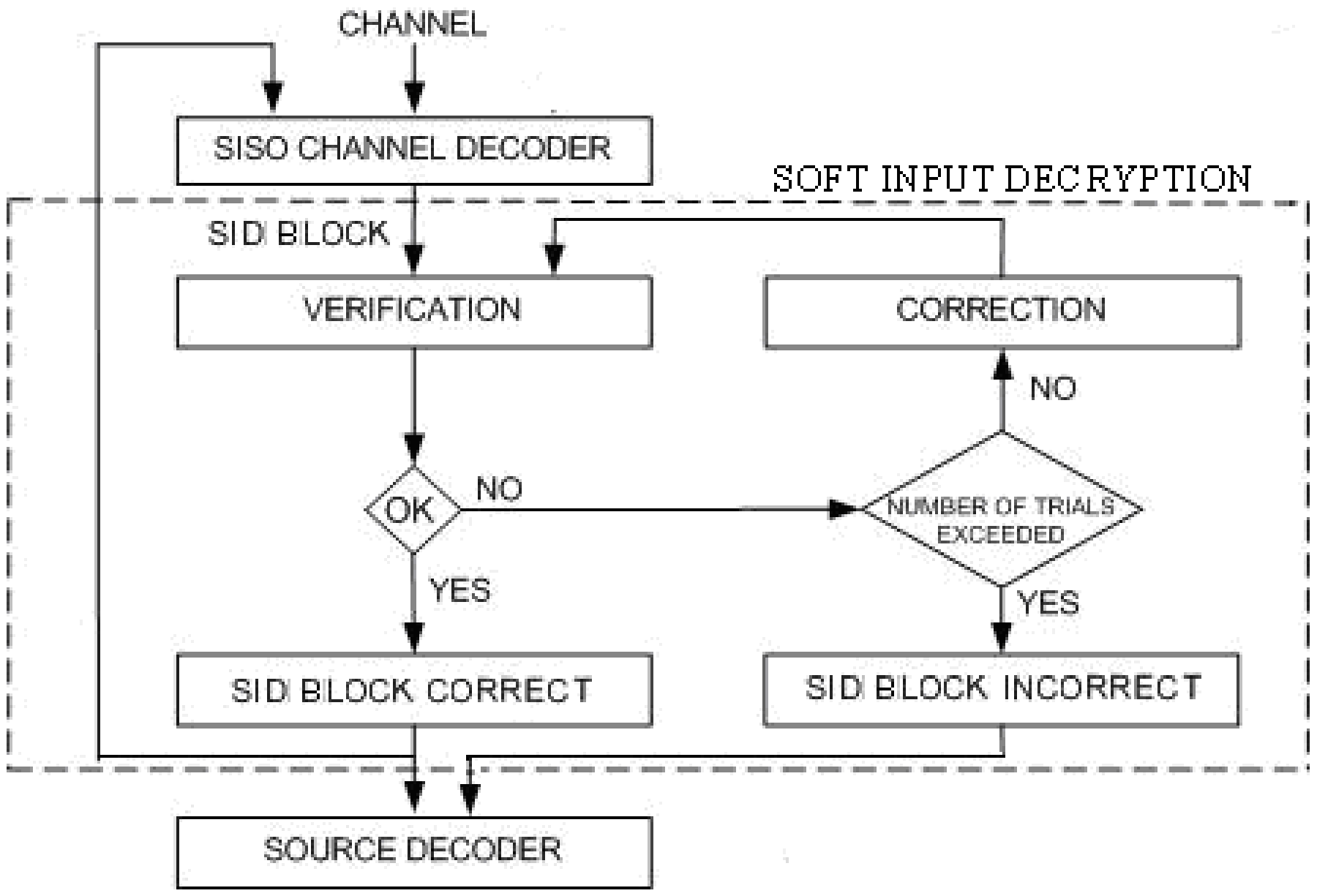

Fig. 4 Soft Input Decryption algorithm

Soft Input Decryption considers the sequence of positions of increasing $|L|$-values. Therefore, at the beginning of the algorithm the bits of the SID block are sorted increasingly according to the $|L|$-values.

If the first verification of cryptographic check values after starting Soft Input Decryption is not successful, the bit with the lowest $|L|$-value of the SID block is flipped, assuming that the wrong bits are probably those with the lowest $|L|$-values (it is also assumed that there are no intentional manipulations). If the verification is again not successful, the bit with the second lowest $|L|$-value is changed. The next try will flip the bits with the lowest and the second lowest $|L|$-value, then the bit with the third lowest $|L|$-value, etc. The process is limited by the number of bits with the lowest $|L|$-values, which should be tested. The strategy follows a representation of an increasing binary counter, whereby the lowest bit corresponds to the bit with the lowest $|L|$-values, and the marked bits correspond to the bits which have to be flipped.

The results of Soft Input Decryption for 320 bit long SID blocks presented in [13] have shown a remarkable coding gain comparing to the results without using Soft Input Decryption. In [13] the coding gain is computed using SER (signature error rate), as a measure for correction of digital signatures. In this paper the coding gain is calculated using BER, as usual in literature and coding theory. This type of realization of the Soft input Decryption algorithm has been used by the simulations of this paper. Other possible realizations, which could speed up or improve the results of testing the $L$- values, are not examined in this paper (for example a L-value group strategy or BER based strategy).

## 4. Feedback

A corrected SID block can be used for the improved error correction of channel decoding of another SID block using feedback [16].

The source encoder outputs data, which are segmented into blocks. Two data blocks are considered as message *ma* (of length $m_1$) and message *mb* (of length $m_2$). Each message is extended by a cryptographic check value *na* (of length $n_1$) rsp. *nb* (of length $n_2$) using a cryptographic check function CCF (Fig. 5), forming blocks *a* and *b*

$$a = a_1 a_2 \ldots a_{m_1+n_1} = ma_1 ma_2 \ldots ma_{m_1} na_1 na_2 \ldots na_{n_1}$$

(1)

$$b = b_1 b_2 \ldots b_{m_2+n_2} = mb_1 mb_2 \ldots mb_{m_2} nb_1 nb_2 \ldots nb_{n_2}$$

(2)

For simplicity and without limitation of generality, it is further assumed that $m_2 \geq m_1$, $n_1 \geq n_2$, and $(m_2 + n_2) \bmod (m_1 + n_1) = 0$. The interleaved blocks *a* and *b* form the joint message *u* (Fig. 5):

$$u = \begin{cases} a_1 b_1 a_2 b_2 \ldots a_{m_1+n_1} b_{m_2+n_2}, \ if\ m_1+n_1 = m_2+n_2 \\ a_1 b_1 \ldots b_{\frac{m_2+n_2}{m_1+n_1}} a_2 \ldots a_{m_1+n_1} b_{m_2+n_2-\frac{m_2+n_2}{m_1+n_1}+1} \ldots b_{m_2+n_2}, if\ m_1+n_1 < m_2+n_2 \end{cases}$$

(3)

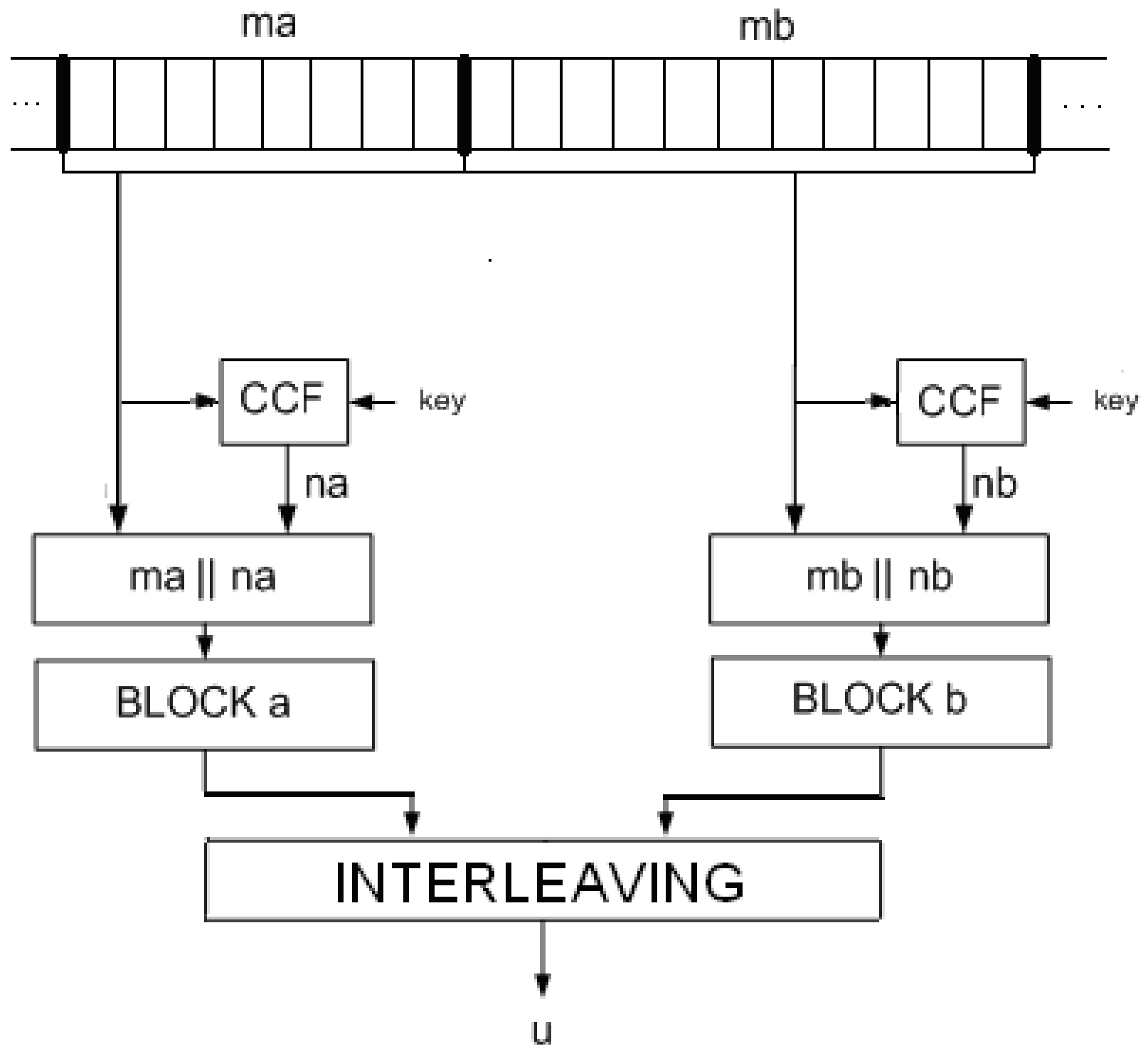

Fig. 5 Formatting message *u*

*u* is encoded, modulated and transferred over an AWGN channel.

The feedback method achieves two steps (Fig. 6): in step 1 the output *u'* of the channel decoder with $BER_{cd1}$ (= BER of the SISO channel decoder) is segmented into blocks *a'* and *b'*, and block *a'* is tried to be corrected by Soft Input Decryption using the *L*-values of *a'*. If Soft Input Decryption is successful, block *a'* is corrected, the *L*-values of the bits of *a'* are set to $\pm\infty$ and the *L*-values of the bits of block *b'* are set to 0. These *L*-values are "fed back" to the channel decoder. In case that Soft Input Decryption is not successful, the 2nd step is skipped and BER remains $BER_{cd1}$. $BER_{1.SID}$ is BER after the 1st step and it is lower than $BER_{cd1}$ because of coding gain introduces by successful Soft Input Decryption of the block *a'*.

In the 2nd step of the feedback method *u'* is decoded again, but now with the fed back *L*-values. The resulted $BER_{cd2}$ is lower than $BER_{cd1}$: the bits of block *a'* are correct and the bits of block *b'* have a lower BER compared to the 1st round of channel decoding. The fact, that $BER_{cd2} < BER_{cd1}$, can be exploited by extension of the length of block b, i.e.: $m_2 + n_2 > m_1 + n_1$.

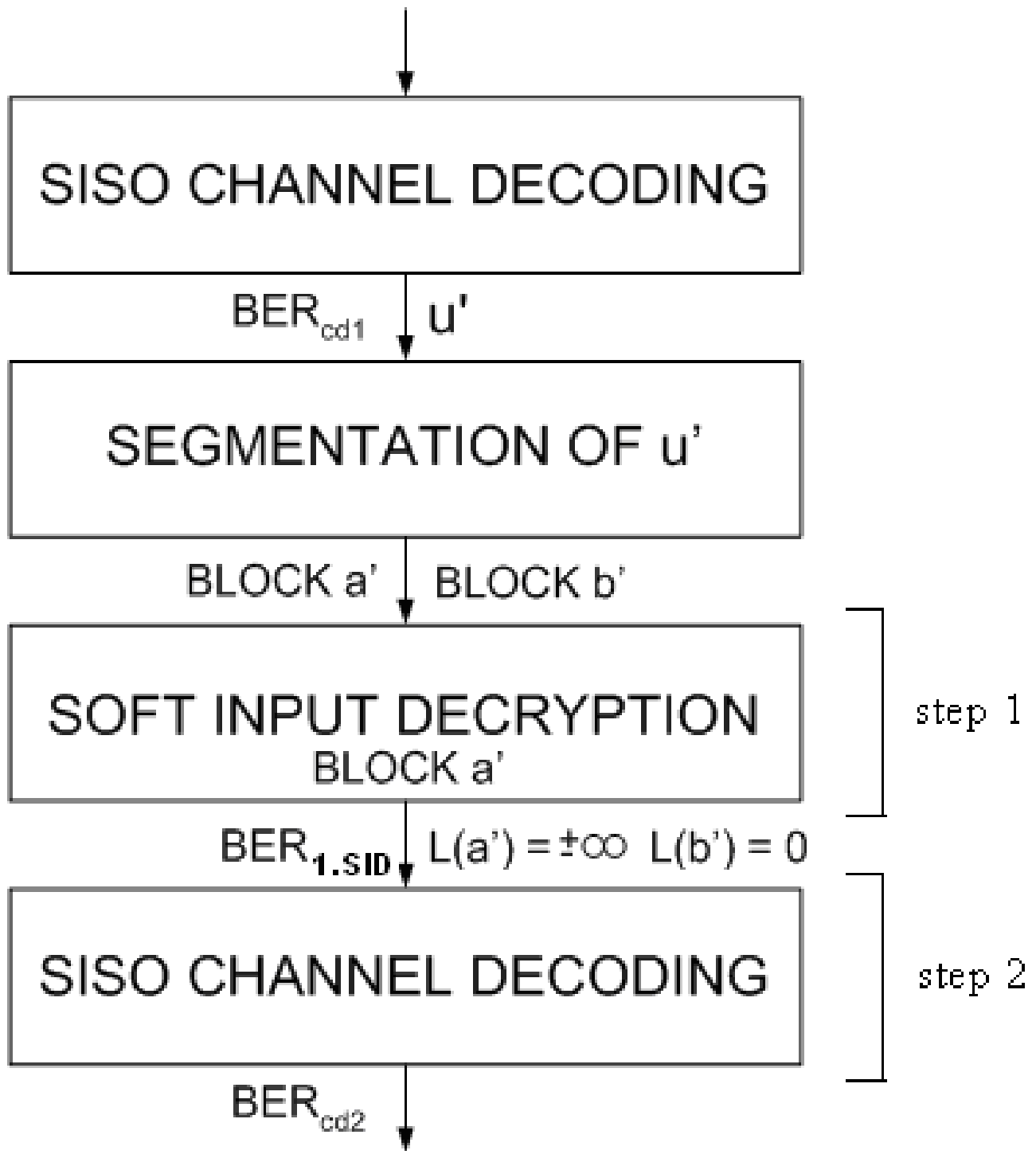

Fig. 6 Algorithm of the feedback method

Simulations are performed by using lengths of block *a* and *b* of 320 bit, i.e. $m_1 + n_1 = m_2 + n_2 = 320$. The individual length of $m_1$ and $n_1$, rsp. $m_2$ and $n_2$ are not important, because they have no impact on *BER* (they have to be considered under security aspects). This corresponds to a signature giving message recovery or with appendix using ECC (Elliptic curve cryptography) over GF($p$) with ld $p$ = 160, rsp. over GF($2^{160}$) (ECNR [17] rsp. ECDSA [4']), or to 256 data bits plus MAC/H-MAC of 64 bit. The transfer is simulated by use of an AWGN channel. The implemented convolutional encoder has a code rate of 1/2 and a constraint length of 2. The decoder uses a MAP [15] algorithm. All simulations are programmed in C/C++ programming language. For each point of the curves 50 000 tests are performed.

The results in [13] have shown that the coding gain of Soft Input Decryption using convolutional and turbo codes is similar. Therefore, only convolutional codes are used in this paper.

The results of the simulations after the 1st and the 2nd step of the feedback algorithm are shown in Fig. 7. For example, for $E_b/N_0$ = 5 dB, $BER_{1.SID}$ is about $10^{-6}$, but $BER_{cd2}$ decreases under $10^{-7}$ with the feedback method.

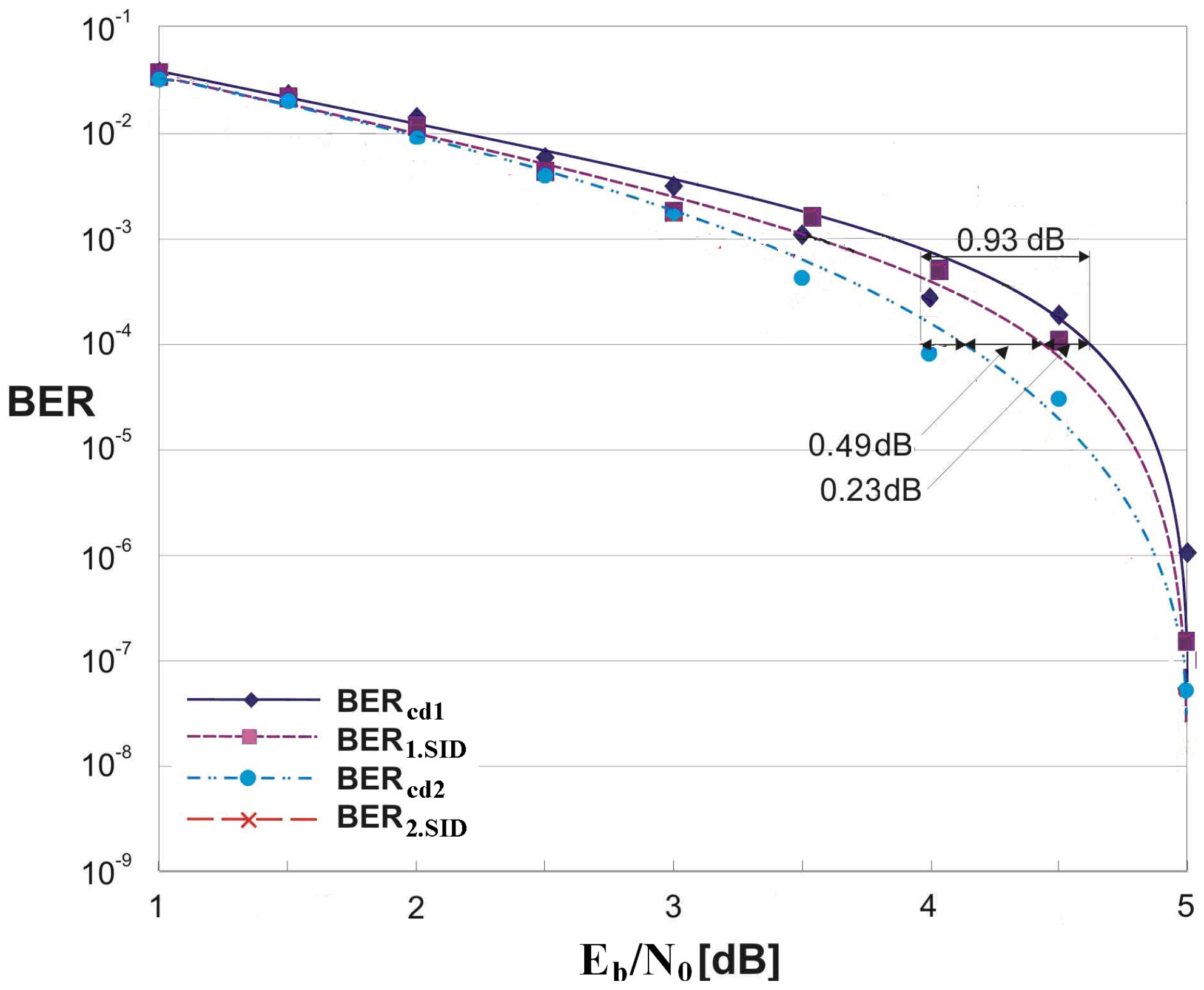

Fig. 7 BER after the 1st and the 2nd step of the feedback algorithm

## 5. Iterative Method

After the 2nd step of Soft Input Decryption with feedback, a 3rd step is introduced - SID of block *b'*.
There are two schemes of Iterative Soft Input Decryption: serial and parallel.

### 5.1. Serial scheme of Iterative Soft Input Decryption

If step 1 - 3 are sequentially performed, the scheme is called serial scheme (Fig. 8).

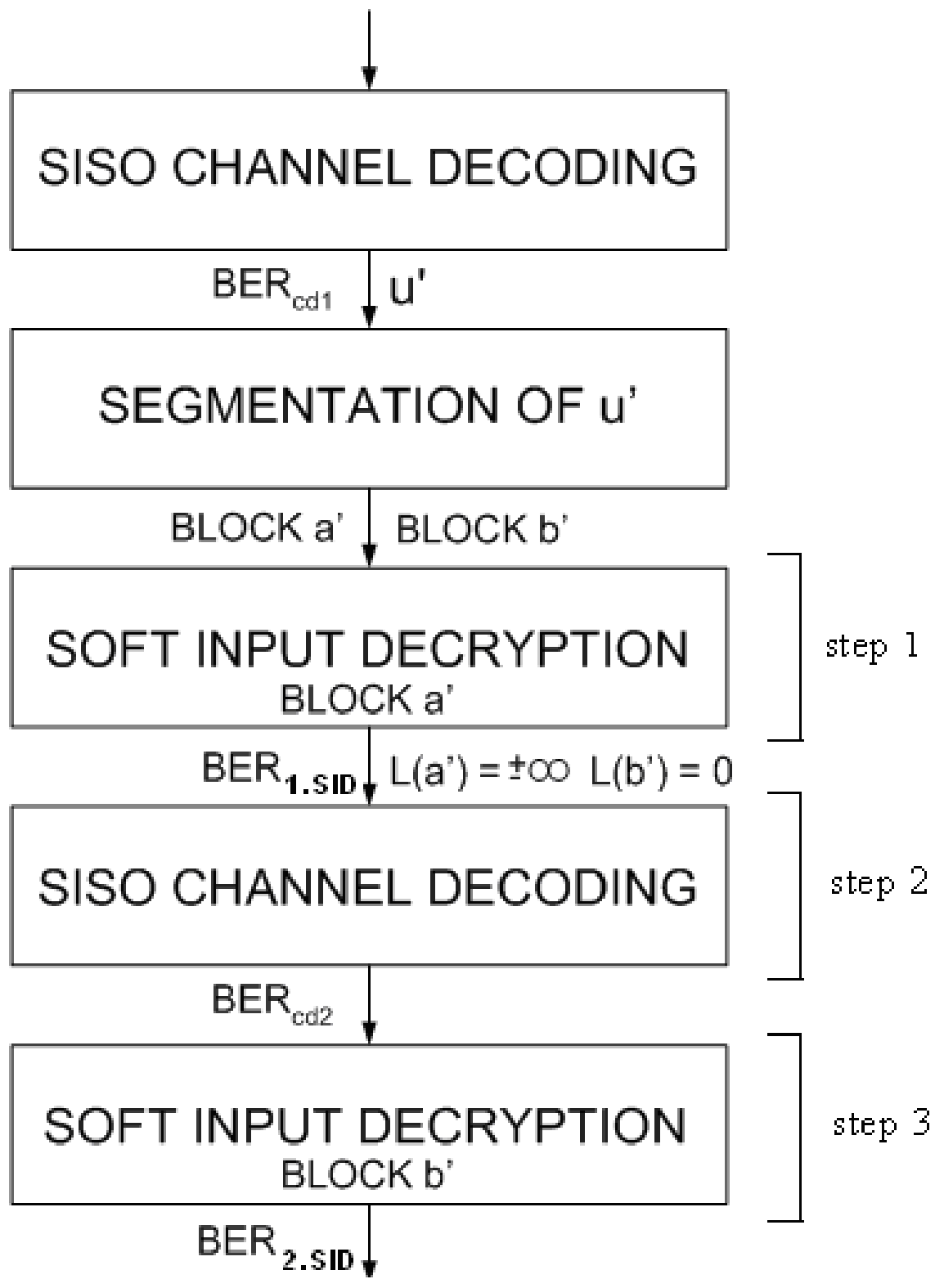

Fig. 8 Algorithm of Serial scheme of Soft Input Decryption with iterations

The simulations use the same parameters as in the simulations before. The length of blocks *a* and *b* is 320 bit. The results of the simulations are presented in Fig. 9. The difference to Fig. 8 is the added function showing BER after the 3rd step ($BER_{cd2}$).
It is obvious, that Soft Input Decryption of block *b* in the 3rd step provides an additional coding gain of about 0.21 dB.

Following simulations of Serial Iterative Soft Input Decryption examine the influence of various lengths of blocks *a* and *b*, but with constant length of *u,* on the coding gain. Coding gains for a length of *u* of 640 bit are shown in Fig. 10 in comparison to channel decoding ($BER_{cd1}$). The message *u* is divided into blocks *a* and *b* of following various lengths:

- test 1:
  block *a* of 128 bit: for example a 64 bit message and 64 bit MAC
  block *b* of 512 bit: for example a 448 bit message and 64 bit MAC
- test 2:
  block *a* of 160 bit: for example a 96 bit message and 64 bit MAC
  block *b* of 480 bit: for example 416 bit message and 64 bit MAC
- test 3:
  block *a* of 212 bit: for example 148 bit message and 64 bit MAC
  block *b* of 428 bit: for example 354 bit of message and 64 bit of MAC
- test 4:
  block *a* of 320 bit: for example digital signatures giving message recovery
  block *b* of 320 bit: for example digital signatures giving message recovery

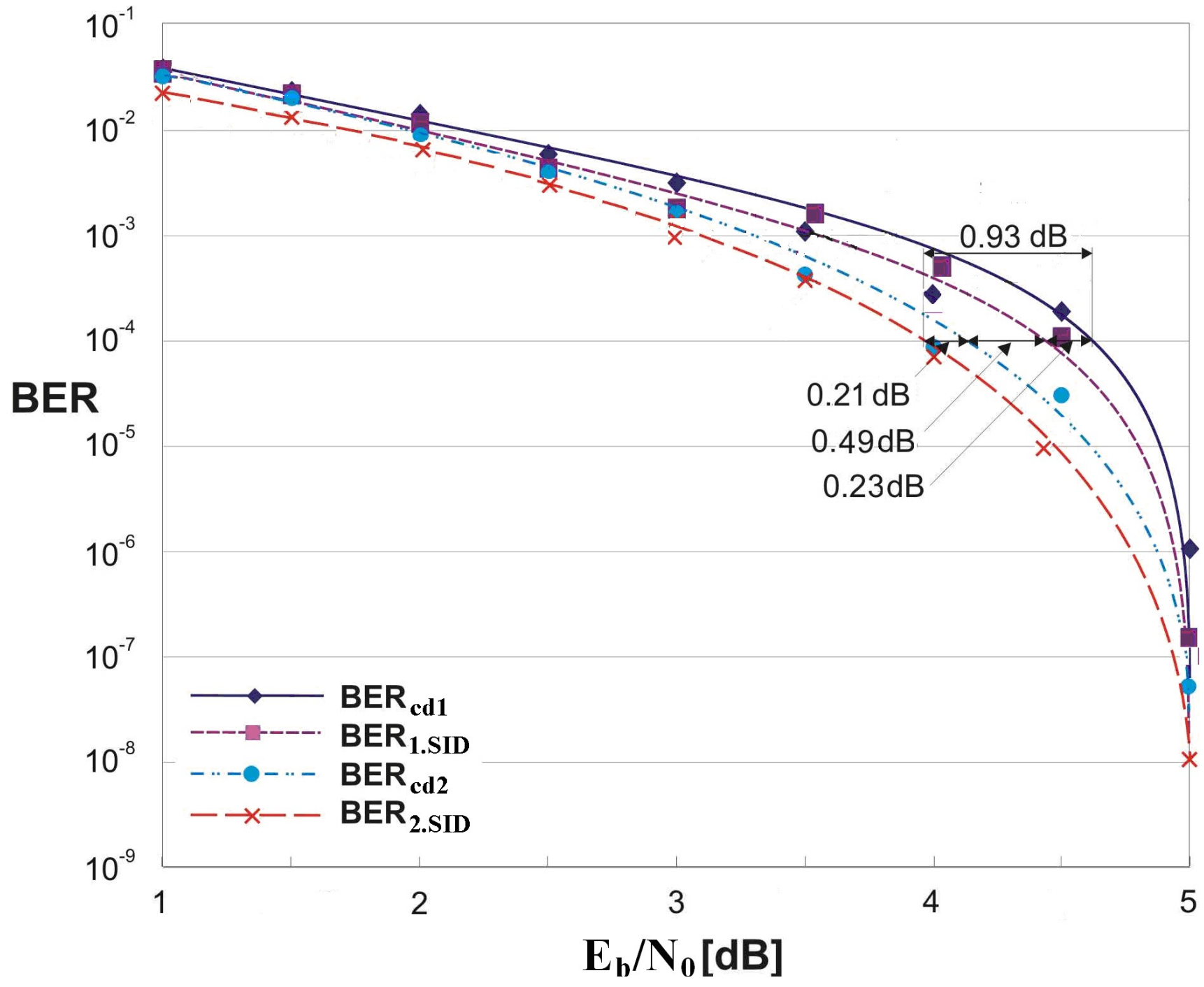

Fig. 9 BER after the 1st , 2nd and 3rd step of the Serial scheme of Iterative Soft Input Decryption

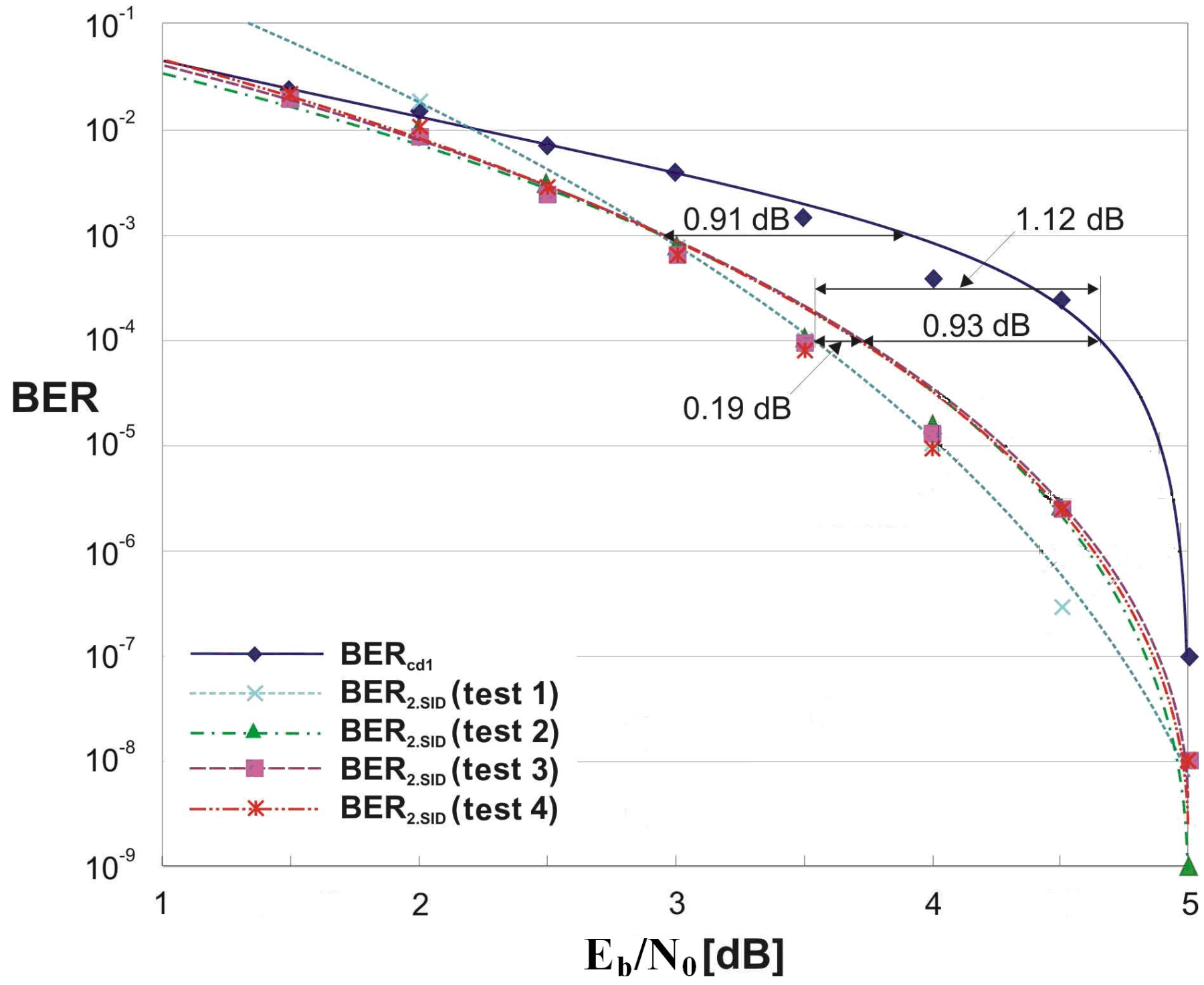

Fig. 10 Coding gains of sequential joint channel decoding and decryption for different lengths of block *a* and block *b* in comparison to channel decoding

The results in Fig. 10 show no significant difference between BER for different lengths of block *a* and *b*. The reason is, that the advantage of a shorter block *a* (better results of Soft Input Decryption) is neutralized by the disadvantage of a longer block *b* (worse results of Soft Input Decryption) and vice versa.

### 5.2. Parallel scheme of Iterative Soft Input Decryption

If the same length is chosen for block *a* and *b*, then it is more efficient to use parallel instead of sequential joint coding and cryptography.

The algorithm of the parallel scheme is as follows: step 1 is performed for block *a´* and *b´* in parallel (Fig. 11). The parallel performance is shown in Fig. 11 using two branches: *a* and *b*. Steps 2 and 3 follow the 1st step in one of the branches, depending on the branch, in which Soft Input Decryption is successful in step 1. In this way Soft Input Decryption with feedback of block *a´* is used for an improved decoding of block *b´* or, vice versa, Soft Input Decryption with feedback of block *b´* is used for improved decoding of block *a´*. After the 1st step is performed in parallel, the 2nd and 3rd step follow in the left or right branch of the scheme (depending on the success of Soft Input Decryption in the left or right branch). The advantage of the parallel scheme in comparison to the serial one is:

1. if the 1st step is successful in both branches, the resulting BER is 0 (i.e. all errors are corrected) and no other steps are performed
2. if the 1st step is not successful in branch *a*, it can be successful in branch *b* and the following steps are performed in branch *b*.

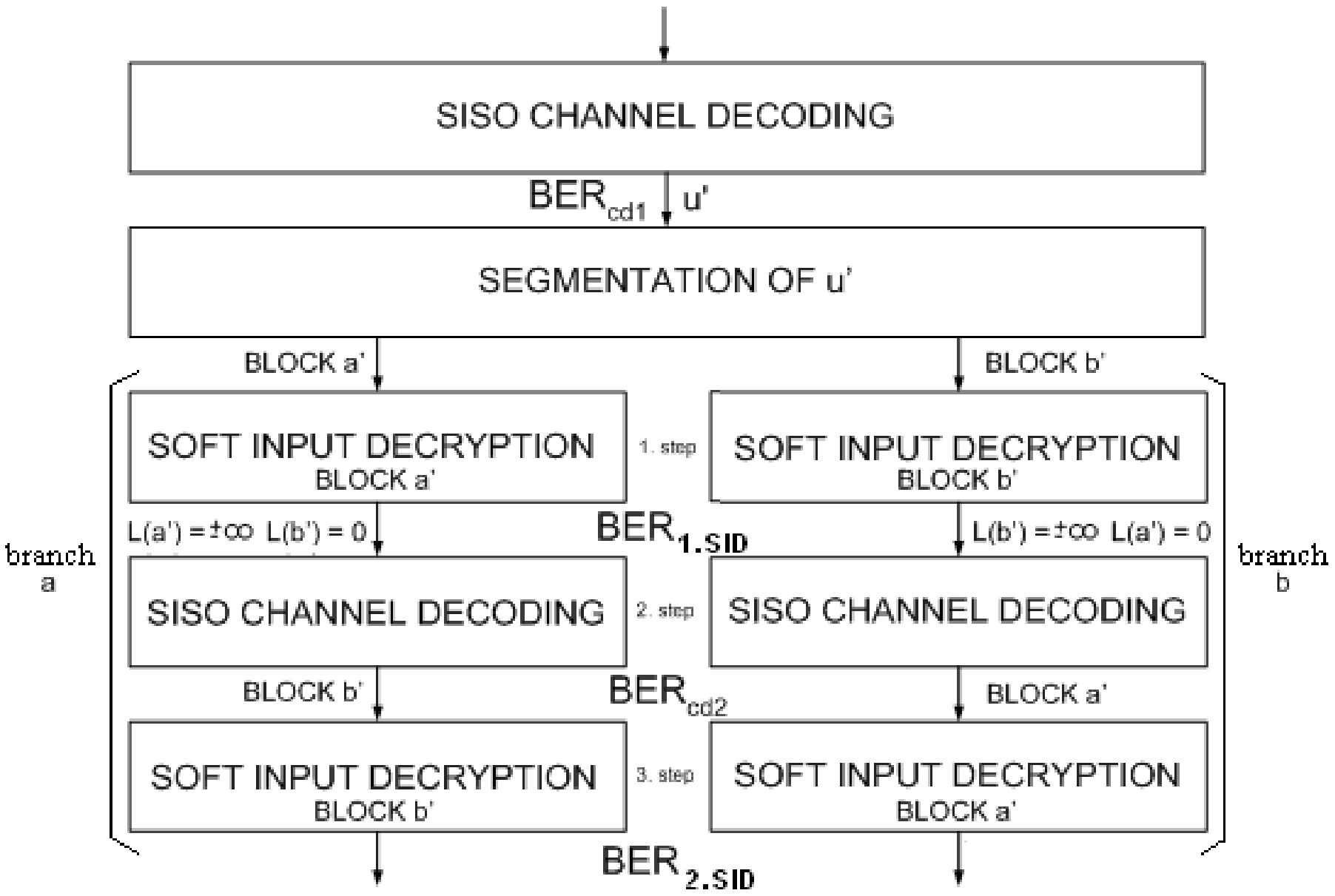

Fig. 11 Parallel scheme of Iterative Soft Input Decryption

Parallel Iterative Soft Input Decryption has been simulated with a length of block *a* and *b* of 320 bit.

The results of the simulations, i.e. BER after each step of the algorithm, are shown in Fig. 12.

After each step BER is calculated as an average BER of branch a and b of the scheme after 50000 tests ("+" means, that in part of cases BER of branch *a* and in other cases BER of branch *b* is calculated):

$$BER_{1.SID} = \text{average value } [(BER_{1.SID} (\text{block } a') + BER_{1.SID} (\text{block } b')] \quad (4)$$

$$BER_{cd2} = \text{average value } [(BER_{cd2} (\text{block } b') + BER_{cd2} (\text{block } a')] \quad (5)$$

$$BER_{2.SID} = \text{average value } [(BER_{2.SID} (\text{block } b') + BER_{2.SID} (\text{block } a')] \quad (6)$$

Fig. 12 shows that the coding gain increases with increasing $E_b/N_0$, because none of the Soft Input Decryption is successful, if too many bits are modified due to a high noise.

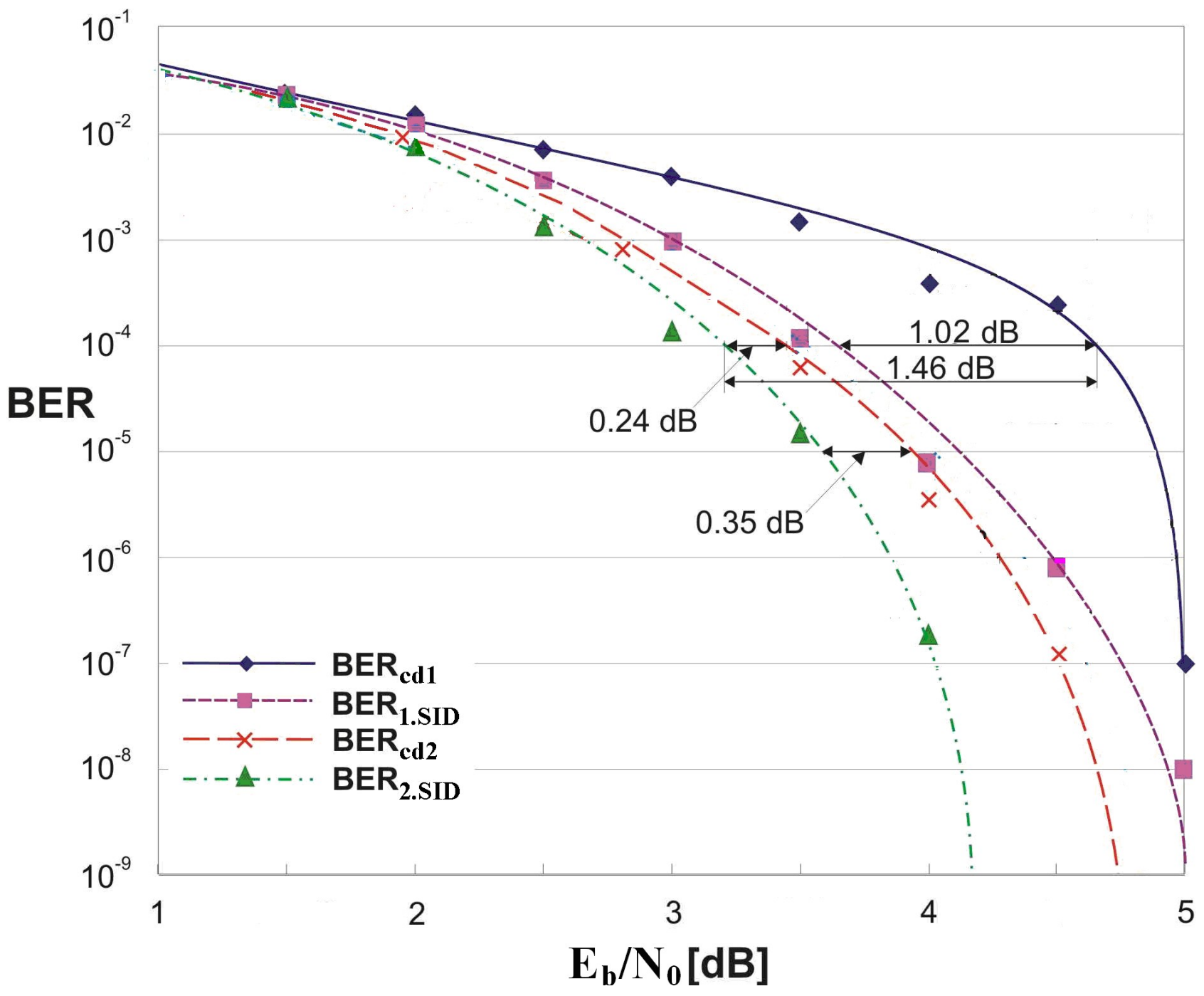

Fig. 12 BER after the 1st , 2nd and 3th step of the Parallel Iterative Soft Input Decryption

Following simulations of Parallel Iterative Soft Input Decryption reflect the influence of various lengths of blocks *a* and *b* on coding gains (blocks *a* and *b* have the same length). Coding gains are shown in Fig. 13 in comparison to $BER_{cd1}$ of channel decoding. Message *u* is divided into block *a* and block *b*, both of them of lengths:

- test 1: of 128 bit: for example 64 bit of message and 64 bit of MAC
- test 2: of 160 bit: for example 32 bit of message and 128 bit of MAC
- test 3: of 256 bit: for example 128 bit of message and 128 bit of MAC
- test 4: of 320 bit: for example digital signatures giving message recovery
- test 5: of 640 bit: for example digital signatures giving message recovery.

The results in Fig. 13 show a significant difference of BER for various lengths of blocks *a* and *b*. The length of *u* influences the results of Soft Input Decryption [18] and therefore also the results of Parallel Iterative Soft Input Decryption.

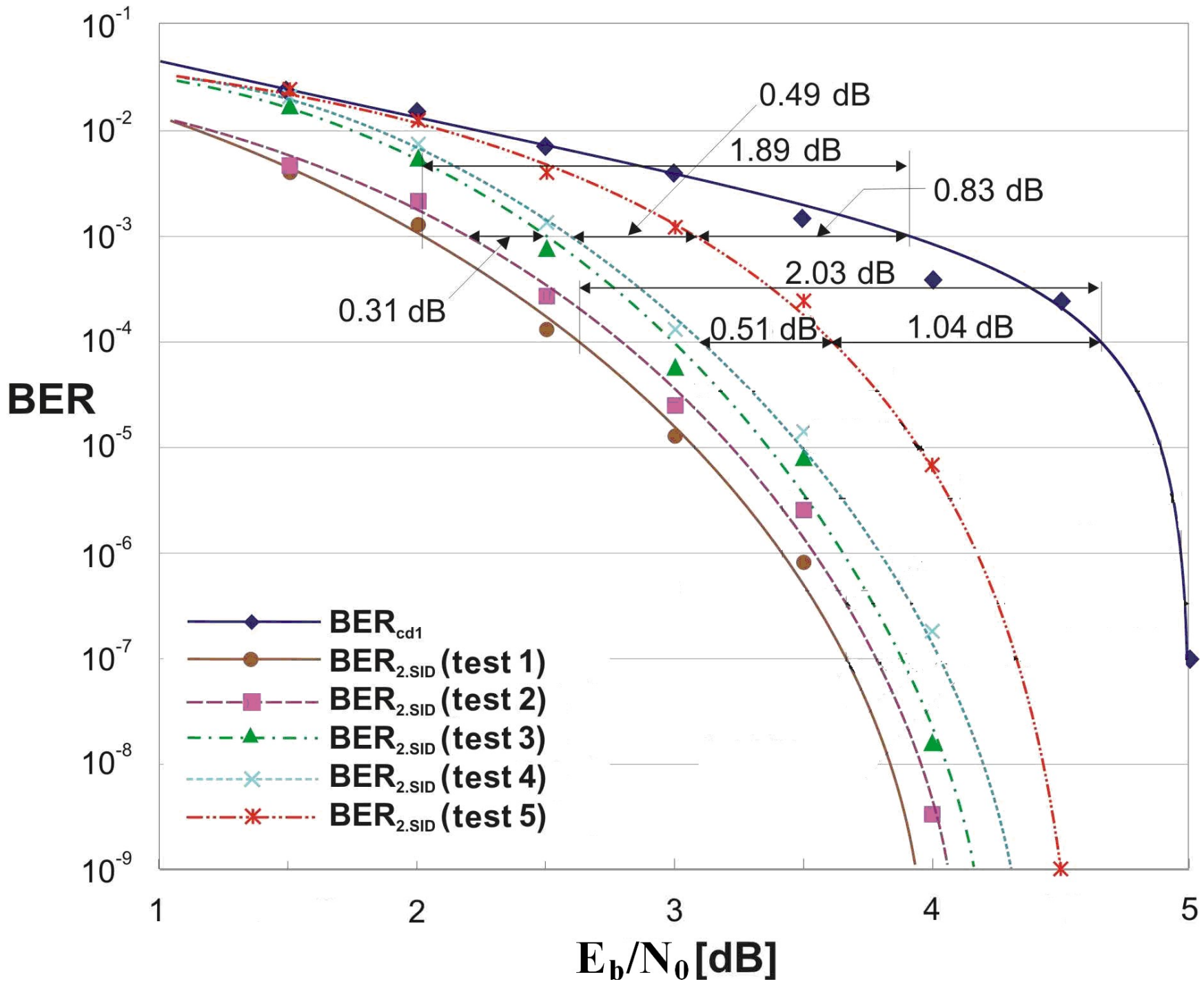

Fig. 13 Coding gains of parallel joint channel decoding and decryption for different lengths of block *a* and block *b* in comparison to channel decoding

### 5.3. Comparison of Serial scheme and Parallel of Soft Input Decryption with iterations

The results after the 3rd step ($BER_{2.SID}$) of the serial and the parallel scheme are presented for a length of *u* of 640 bit in Fig. 14: blocks *a* and *b* have the same length of 320 bit. Coding gains are obtained by comparison to $BER_{1.cd}$. The results show a big difference between the parallel and serial scheme: the coding gain of a parallel scheme is up to 0.82 dB higher than the coding gain of a serial scheme.
For that reason the parallel scheme is recommended to be used. As both SID blocks have the same length, parallel scheme is easier for implementation (easier segmentation). Also, the same lengths of SID blocks imply the same security level and collision probability of SID blocks during verification.

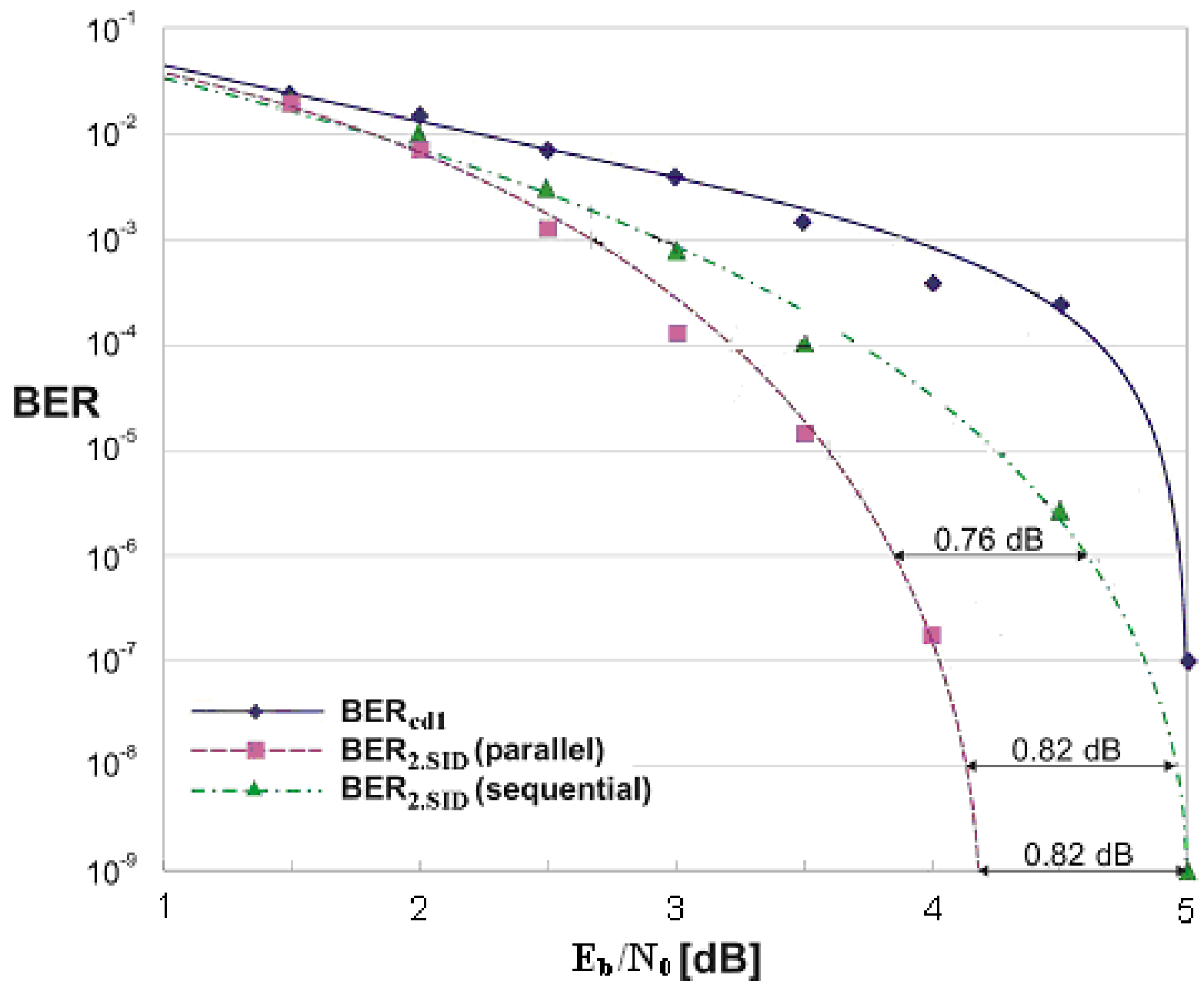

Fig. 14 Comparison of coding gains of Serial and Parallel Soft Input Decryption with iterations

## 6. Conclusion and Future Work

Error sensitivity of digital signatures and other cryptographic mechanisms call for correct cryptographic information for which the Soft Input Decryption is a very good approach, especially in environments with a low signal-noise ratio. Soft Input Decryption uses a channel code plus cryptographic check values as redundancy for improved decoding.
Methods of Soft Input Decryption and Iterative Soft Input Decryption are described in this paper and results of simulations are presented. The results of the simulations show a significant coding gain of both methods compared to the case when the methods are not used. The coding gain of Iterative Soft Input Decryption of two SID blocks of length of 320 bit reaches 1.2 dB.

If the cryptographic check values are used by security needs, the improvement of decryption and channel decoding is free of cost. Under coding aspects the coding gain is paid by a lower code rate. It is not aimed, that the use of cryptographic check values is the best way of improvement of channel coding just under channel coding aspects if the reduced code rate is considered.

Iterative Soft Input Decryption can be extended to more than one iteration. In such a case, more than two SID blocks can be iteratively decoded for getting a higher coding (turbo principle). Analysis and simulations of a higher number of iterations are a suggestion for future work.

**Authors**

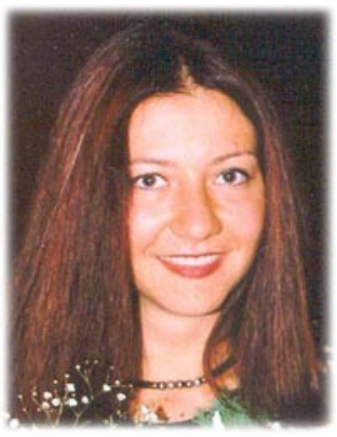

**Natasa Zivic, Dr.,** born 1975 in Belgrade, Serbia, graduated from the Faculty of Electrical Engineering (Electronics, Telecommunication and Automatics) of the Belgrade University in 1999. at the Telecommunication Department. After the Post diploma studies at the same Faculty (Telecommunications Division) she defended her Magister Thesis (Acoustics) in 2002.
From October 2004. she was scientific assistant at the University of Siegen in Germany at the Institute for Data Communications Systems as a DAAD and University of Siegen Scholarship holder. In 2007. she defended her Doctoral Thesis on the same University. The main course of her work in Siegen is Coding and Cryptography. From 2000. till 2004. she was working at the Public Enterprise of PTT "Serbia", Belgrade as the senior engineer. Currently she is employed as an Assistant Professor at the University of Siegen.